\documentclass[preprint]{vgtc}               

\graphicspath{{figures/}{pictures/}{images/}{./}} 

\usepackage{times}                     

\usepackage{tabu}                      
\usepackage{booktabs}                  
\usepackage{lipsum}                    
\usepackage{mwe}                       

\usepackage{mathptmx}                  
\usepackage{ccicons}
\usepackage{xfrac}
\usepackage{musicography}

\newcommand{\new}[1]{{#1}}

\newcommand{\osflink}[1]{\href{https://osf.io/zx3ac}{#1}}
\newcommand{\osflinkrepo}{\osflink{\texttt{osf\discretionary{}{.}{.}io\discretionary{/}{}{/}zx3ac}}\xspace}

\onlineid{0}

\vgtccategory{Research}

\vgtcinsertpkg

\preprinttext{This manuscript was presented at alt.VIS, a workshop co-located with IEEE VIS 2025.}

\title{Data Melodification FM: Where Musical Rhetoric Meets Sonification}

\author{Ke Er Amy Zhang\thanks{e-mail: ke.zhang@uib.no}\\ %
        \scriptsize University of Bergen %
\and David Grellscheid\thanks{e-mail: david.grellscheid@uib.no}\\ %
     \scriptsize University of Bergen %
\and Laura Garrison\thanks{e-mail: laura.garrison@uib.no}\\ %
     \scriptsize University of Bergen}

\teaser{
  \centering
  \includegraphics[width=0.85\linewidth]{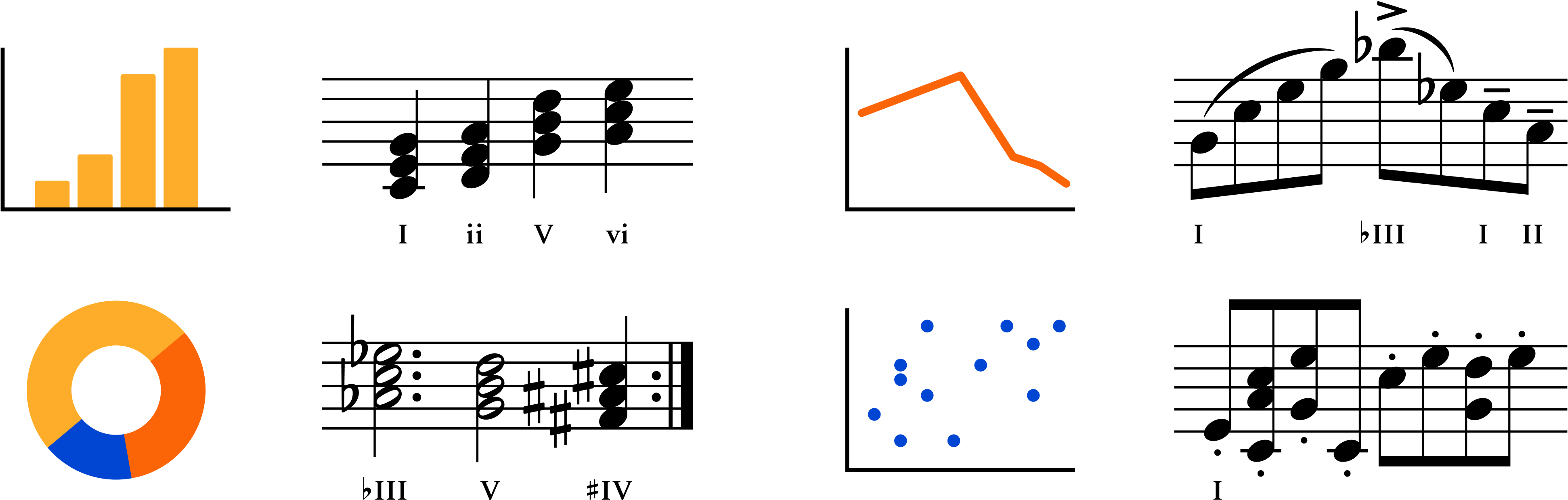}
  \caption{Conceptual overview of visualization idioms and their \textit{melodification} counterparts.}
  \label{fig:teaser}
}

\abstract{
    We propose a design space for data \textit{melodification}, where standard visualization idioms and fundamental data characteristics map to rhetorical devices of music for a more affective experience of data. Traditional data sonification transforms data into sound by mapping it to different parameters such as pitch, volume, and duration. Often and regrettably, this mapping leaves behind melody, harmony, rhythm and other musical devices that compose the centuries-long persuasive and expressive power of music. What results is the occasional, unintentional sense of tinnitus and horror film-like impending doom caused by a disconnect between the semantics of data and sound. Through this work we ask, can the aestheticization of sonification through (classical) music theory make data simultaneously accessible, meaningful, and pleasing to one's ears?
} 


\begin{document}


\firstsection{Ouverture}
\maketitle

Sonification is a well-established approach for mapping data to sound for accessibility purposes as well as for broader public outreach. Rather than positioning data in physical/pixel space, data sonification positions data as auditory marks in sequence, i.e., data points are played as individual sounds over a period of time. These marks are encoded using auditory channels that include pitch, volume, tempo, and timbre~\cite{wang2022seeing}. For instance, volume, pitch, and tempo may be leveraged together to encode the arrivals and departures of different bus lines at a given stop~\cite{ronnberg2024visualization}. These channels vary in their effectiveness, even with following current suggestions for best practices for natural, task-appropriate data-to-sound mappings~\cite{gross2023reflecting}.
Implemented in tools like the \textit{HighCharts Accessibility Module and Sonification Studio}\footnote{sonification.highcharts.com/\#/app}, these auditory channels enable users with or without visual impairments to create, explore, and share data visualizations that can be both seen and heard~\cite{enge2024open}. In outreach, NASA has translated astronomical data to sound to allow an audience to ``hear'' a black hole, instilling a sense of wonder while correcting the pervasive misconception that all of space is a soundless vacuum~\cite{nasa2022sonificationblackhole}. Data sonification, if crafted with careful intention, additionally shows promise in conveying emotion alongside information, although this comes with some precision cost~\cite{ronnberg2021sonification}. 

Regrettably, a majority of data sonifications do not yield a particularly aesthetic experience. As sighted users, we authors found the auditory experience of several data sonifications to signal through, e.g., increasing pitch and volume, a sense of impending doom or, at the very least, make us feel a bit more twitchy than expected. Through this work, we aim to create a more aesthetic data sonification experience inspired by melodies in classical music---in other words, data \textit{melodification}. 

We are not the first people interested in more aesthetic, less dread-inducing ways to sonify data. Drawing from the natural world, Hoque et al.'s~\cite{hoque2023accessible} design space of rain showers, bird calls, and other nature sounds create a soothing data soundbath. Our design space similarly considers the pleasing aspects of sonification for basic idioms, but draws instead from classical music. Barrass~\cite{barrass2012aesthetic} argues for an ``aesthetic turn'' in data sonification that integrates art with science, and functionality with aesthetics in the \textit{TaDa} method. We build on this notion in our design space to collapse so-called \textit{mechanistic} scientific with \textit{humanistic} artistic methods. 

\begin{figure}[tb]
 \centering 
 \includegraphics[width=\linewidth]{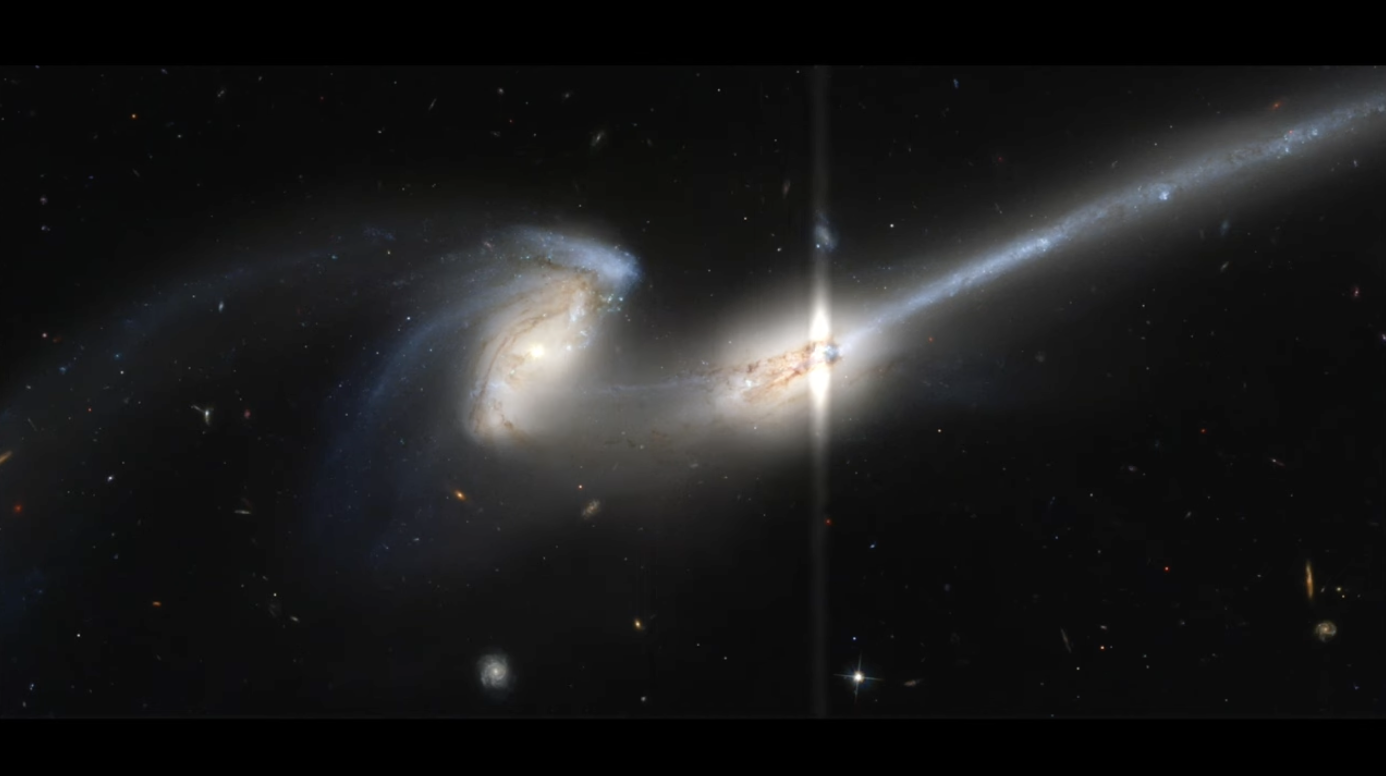}
 \caption{Sonification of the \href{https://youtu.be/KzwBOJi0PA4}{Mice Galaxies} image, taken by the NASA Hubble Space Telescope. Playhead over right galaxy. Public domain.}
 \label{fig:galaxies}
\end{figure}

Data--sound aesthetic mapping forms a natural continuum with sound art. The region of this continuum \new{that leans towards} sound art while retaining sufficient remnants of the data to communicate a message is useful for scientific communication and outreach, as explored by the \textit{Air Listening Station} project~\cite{larrieux2023air}. This project transforms air quality measurements to immersive soundscapes through generative music and algorithmic composition. Their design techniques informed our approach to crafting our design space, particularly in our initial experiments with live-coding tools. Similar public science communication projects that reimagine data as aesthetic, emotional experiences include NASA's orchestral soundscapes of Hubble Telescope images~\cite{nasa2025sonificationhubble}. Here, \new{NASA scientists collaborate with music composers} to encode visual channels such as brightness with pitch and volume, while vertical position further modulates the pitch of stringed instruments played by human musicians. \new{We link the  \href{https://youtu.be/KzwBOJi0PA4}{Mice Galaxies} shown in \cref{fig:galaxies} as an audio example of this composition process and similarly hyperlink other sonic experiences throughout this manuscript.} Our design space introduces a similar sense of linear temporality with classical music sounds, although rather than image data we focus on \textit{melodifying} visualizations generated through tabular data. The \textit{Leander} project expands from a single to stack of images to generate soundscapes---by varying pitch, rhythm, timbre, bells, and pads---from video by sonifying pixel colors~\cite{hall2020leander}. Puca-Mendez and Wells~\cite{pucamendezMusicScore} describe the converse of what previous works and we attempt here: they convert sheet music to an array, which then maps to channels such as position and color. 

Also related to our project is the \textit{musification} approach for time-series data~\cite{globXblog2020waggawagga}, which maps, e.g., precipitation to pitch constrained within the note range of an \href{https://vimeo.com/411936569?p=1s}{A-minor pentatonic scale and accompanied with guitar chords and base line} in the sonic equivalent of visual embellishments a.k.a. ``chart junk''. Our design space similarly employs a scale-based tonal range to melodically map data values, but extends beyond time-series data. The \textit{Xenakis} project~\cite{schetinger2021xenakis} similarly explores musification paired with visualization, though focusing on topological data to convey the sense of an urban landscape. 

Building upon these aesthetic data works, we are equally inspired by the relationship between mathematics and composed music, as was Bach through his masterful counterpoint with intricate structures and symmetries~\cite{kempf1996symmetry}, Xenakis through his stochastic compositions derived from probability theory~\cite{gibson2011instrumental}, or simply through viral \textit{\href{https://www.youtube.com/shorts/H-JB8JM9R0g}{Will it Riff?}}~\cite{small2024riff} challenges converting barcode digits into guitar tabs. Our curiosity (and fear of some sonification practices) have galvanized us to formally \textbf{map the rhetorical layer of visualization to rhetorical devices in classical music}. Inspired by Blair et al.'s~\cite{blair2024quantifying} work on affective visualization characteristics, we explore how visualization idioms, colour palette, and certain immutable data characteristics can be transformed into sounds of music.

\section{Music Theory Crash Course}

But how does one \textit{hear} data melodification? We give a crash course on (classical) music theory to help with reading and understanding the musical devices included in our design space. 

\begin{figure}[tb]
 \centering 
 \includegraphics[width=\linewidth]{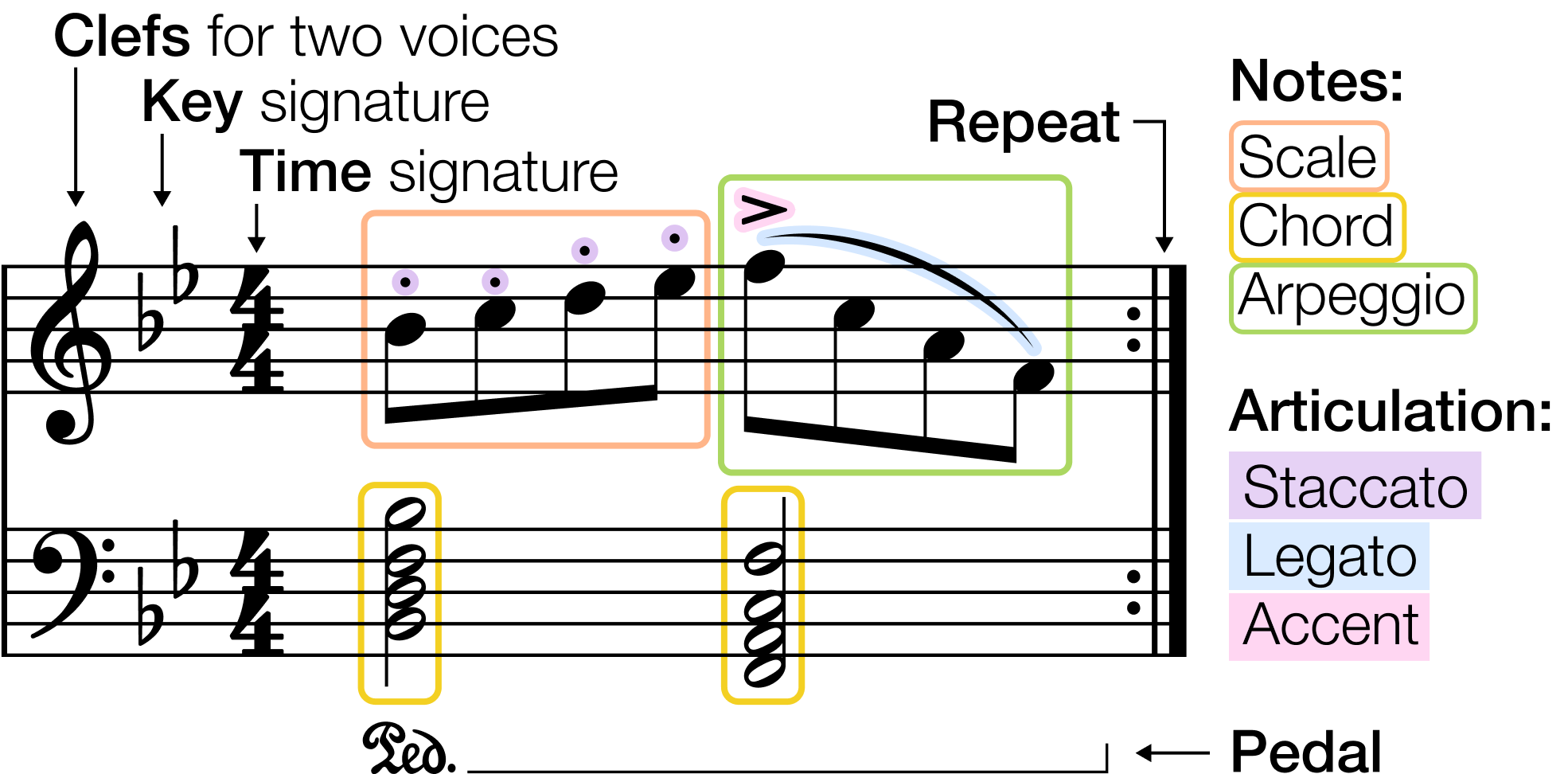}
 \caption{Annotated musical notation \new{for a bar of music.}}
 \label{fig:notation}
\end{figure}

\textbf{Fundamental note characteristics---}Musical notation is the visual language of music---a set of instructions composers write to communicate how a musical piece should be played. Modern notation features five horizontal lines on which \textbf{notes} are placed to indicate the pitch and duration of sound (see~\cref{fig:notation}). In Western music, these notes are drawn from 12 distinct pitches and anchored in ranges specified by the \textbf{clef}; this often sets apart different voices in a piece. To further sustain the duration of a note and reverberate it, a \textbf{pedal} can be held as the note is played. Notes of various durations can be combined into beat patterns \new{within a segment of music called a bar;} this is dictated by a \textbf{time signature} and altogether creates musical \textbf{rhythm}. In our design space, we assign these basic note characteristics to immutable data characteristics, such as pitch to trends, rhythm and reverb to density, and range to variance.

\textbf{Rhetorical layer of music---}Music gains rhetorical power through structured patterns and relationships between its fundamental characteristics. \textbf{Scales}, which are ordered sequences of notes, offer a tonal ``color palette'' that forms the foundation of \textbf{chords}---vertical groups of notes that are played simultaneously or broken apart as \textbf{arpeggios}---as well as \textbf{melodies}---linear patterns of notes that form musical themes. 
These patterns can be built using all 12 notes, e.g., chromatic scale, or built around a \textbf{key signature} that dictates the central pitch of the piece, e.g., G-major chord or E-minor arpeggio. \textbf{Articulations} increase the expressiveness of notes by specifying the length, volume, and attack of individual notes, e.g., \textbf{staccato} for short and detached notes, \textbf{legato} for smooth and connected notes, and \textbf{accents} for strongly emphasized notes. By structuring and expressing notes in specific ways, composers can form recognizable \textbf{melodic patterns}, played \textbf{on repeat} over an underlying \textbf{harmony} created by moving chords in progression. In our design space, we use the recognizable quality of melodic patterns to distinguish between categorical and quantitative data patterns captured in standard visualization idioms.

Often, these melodic and harmonic choices follow a set of stylistic conventions, or \textit{musical rhetoric}, that creates different emotional experiences for the listener. We employ such stylistic conventions to signify the colour palette of a visualization which similarly elicits emotional experiences in an audience~\cite{blair2024quantifying}. In classical music, \textbf{time signature} can contribute to the overall character of a musical piece, e.g., \musMeter{3}{4} time with three beats per bar is associated with waltzes, while \musMeter{2}{4} time with two beats per bar is used for marches.  \textbf{Key signatures} are also associated with distinct affective qualities~\cite{mccreless2002rhetoric}, e.g., the key G-major is seen as bright and innocent, e.g., \href{https://commons.wikimedia.org/wiki/File:Mozart_-_Eine_kleine_Nachtmusik_-_1._Allegro.ogg}{Mozart's \textit{Eine kleine Nachtmusik}}, while D-minor dark and melancholic, e.g., \href{https://commons.wikimedia.org/wiki/File:Prelude_Op._28_no._24.mp3}{Chopin's \textit{Prelude No. 24 ``The Storm''}}. Harmony created by melodic progressions can also convey meaning, as illustrated by counterpoint theory~\cite{schubert2002counterpoint} \new{(famously mastered by Bach)} which specifies rules of harmony for interweaving independent melodies. Individual notes in different voices should not be played simultaneously if they form \textit{``forbidden''} distances or \textbf{intervals} between notes that create dissonance, e.g., \href{https://www.npr.org/2017/10/31/560843189/the-unsettling-sound-of-tritones-the-devils-interval}{the \textit{devil's interval}}~\cite{npr2017tritone}. In a similar vein, we reduce dissonance and feelings of dread in our design space by drawing pitches from standard scales in classical music, e.g., using 12 pitches rather than all possible pitches. In counterpoint theory, musical sections should also be ended and resolved with chord progressions known as \textbf{cadences}, with some cadences giving a feeling of complete resolution, e.g., \href{https://commons.wikimedia.org/wiki/File:Perfect_cadence.mid}{perfect cadence}, while others creating uncertainty, e.g., \href{https://commons.wikimedia.org/wiki/File:Deceptive_cadence_in_C.mid}{deceptive cadence}. We consider the affective influence of these cadences when composing our design space to avoid creating unintentional semantics.
%


\begin{figure*}[ht]
 \centering 
 \includegraphics[width=\linewidth]{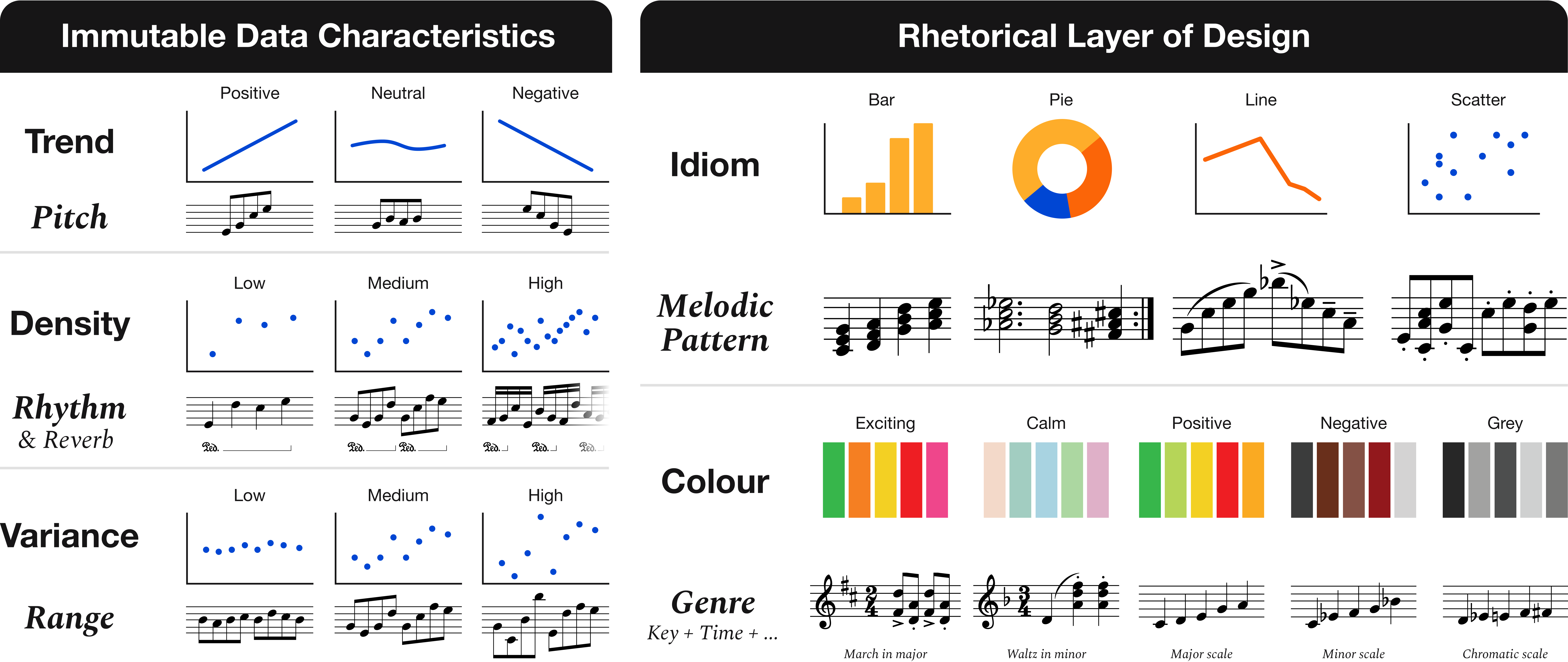}
 \caption{The data melodification design space connects musical devices to immutable data characteristics and rhetorical layers of design.}
 \label{fig:designspace}
\end{figure*}

Though the examples in this manuscript focus on classical music, we imagine the design space of data melodification can be expanded to other styles of music to create even stronger semantic associations. \textbf{Genre}---which can be thought of as distinct combinations of keys, time signatures, and instrumental voices---is able to associate musical structure with time, geography, and more broadly, culture~\cite{lena2008classification}. Whilst classical symphonies are often associated with tradition and formality in pre-20th century Europe, swing jazz in 1940s America is associated in contrast with liveliness, spontaneity, and pure fun. Writing from Bergen, we cannot fail to mention 1980s Norwegian black metal that opposes classical conventions through liberal use of the dissonant \textit{devil's interval} (and yes, this might have Bach rolling in his grave).

\section{Sound design space}
\label{sec:design_space}

While current practices in data sonification primarily focus on mapping musical parameters to visual marks and channels, we instead focus on a higher level of abstraction. Over multiple jam sessions on Sonic Pi and the physical piano, composers Zhang, Grellscheid, and Garrison crafted a design space for melodifcation (see~\cref{fig:designspace}) that connects the rhetorical analogues of music to visualization idioms and colour palettes as well as immutable data characteristics of data trend, density, and variance as described by Blair et al.~\cite{blair2024quantifying}.

\textbf{Immutable data characteristics---}We map \textbf{data trends} to \textbf{pitches of notes}, where a positive trend with ascending data values corresponds to ascending notes, a negative trend to descending notes, and a neutral trend to repeated notes that are close to one another in pitch. We then map \textbf{data density} to \textbf{rhythm}, where increased data density corresponds to an increased density of notes in a segment of music, i.e., more division of beats. We also map density to \textbf{reverb} created through pedalling, where decreased data density corresponds to more reverberations, much like how sound echoes in a large and empty room. Lastly, we map \textbf{data variance} to \textbf{pitch range}, where a wider spread of data points corresponds to a wider frequency of pitches/notes available to be played, e.g., 36 notes rather than 12 notes. Our mappings aligns with current sonification practices for the most part, though a primary difference in our design space is that we take pitches from the standard 12-note system in classical music rather than from all possible values. 

\textbf{Rhetorical layer of design---}We focus on mapping categorical and quantitative data patterns to particular \textbf{melodic patterns}, using the design of four basic \textbf{visualization idioms}---bar, pie, line, and scatter---as inspiration but also a "sanity test" for our melodification decisions. We play categorical data, often represented as vertical columns in a \textbf{bar chart}, as vertical \textbf{chords}, where the progression of chords correspond to the relative differences in value between bars. To represent categorical data as proportions of a whole, we use the duration of the chord as an additional channel that corresponds to the ratio of a category to the whole. The visual equivalent would be a radial pie chart, as such we are inspired to play this categorical data pattern ``circularly'' \textbf{on repeat}. In contrast, we play quantitative data as a different melodic pattern---detached \textbf{staccato} notes within a range determined by the data variance to give a sense of the separated and disconnected data points, equivalent to a \textbf{scatterplot}. For quantitative data in continuous intervals, we play the notes as \textbf{arpeggios} in a \textbf{legato} style---to reflect the smooth and linear connections between data points, equivalent to \textbf{line charts}. The progression of arpeggios corresponds to the relative differences in the slope of the trend lines while the duration of the arpeggio is dependent on the length of the trend line. When the trend changes, we signal this change by accenting the first note of the new arpeggio. 

We map \textbf{colour palette} to \textbf{musical genre} which is often dictated by key and time signature along with instrumental voices. We connect these rhetorical encoding devices as both have influences over the emotional experience of the audience~\cite{blair2024quantifying,gabrielsson2016expression}. Here we describe a selection of colour palettes from Blair et al.~\cite{blair2024quantifying} and what we see as an example of their associated tonal palette. An exciting colour palette is akin to an exuberant and grand orchestral piece best embodied by \href{https://commons.wikimedia.org/wiki/File:Gustav_Holst_-_the_planets,_op._32_-_iv._jupiter,_the_bringer_of_jollity.ogg}{Holst's \textit{Jupiter, Bringer of Jollity} from The \textit{Planets}}, while a calm palette reflects a meditative and repetitive piece, such as \href{https://commons.wikimedia.org/wiki/File:Gymnopedie_No._1..ogg}{Satie's \textit{Gymnop\'edies}}. We can also leverage the affective qualities associated with key signatures, matching palettes that create positive affect with \href{https://commons.wikimedia.org/wiki/File:PentMajor.mid}{major scales}, negative with \href{https://commons.wikimedia.org/wiki/File:C_minor_pentatonic_scale.mid}{minor scales}, and neutral/grey with \href{https://commons.wikimedia.org/wiki/File:Chromatic_scale_ascending_on_C.mid}{chromatic scales}.

\section{Live from the recording studio}
\begin{figure}[htb]
 \centering 
 \includegraphics[width=\linewidth]{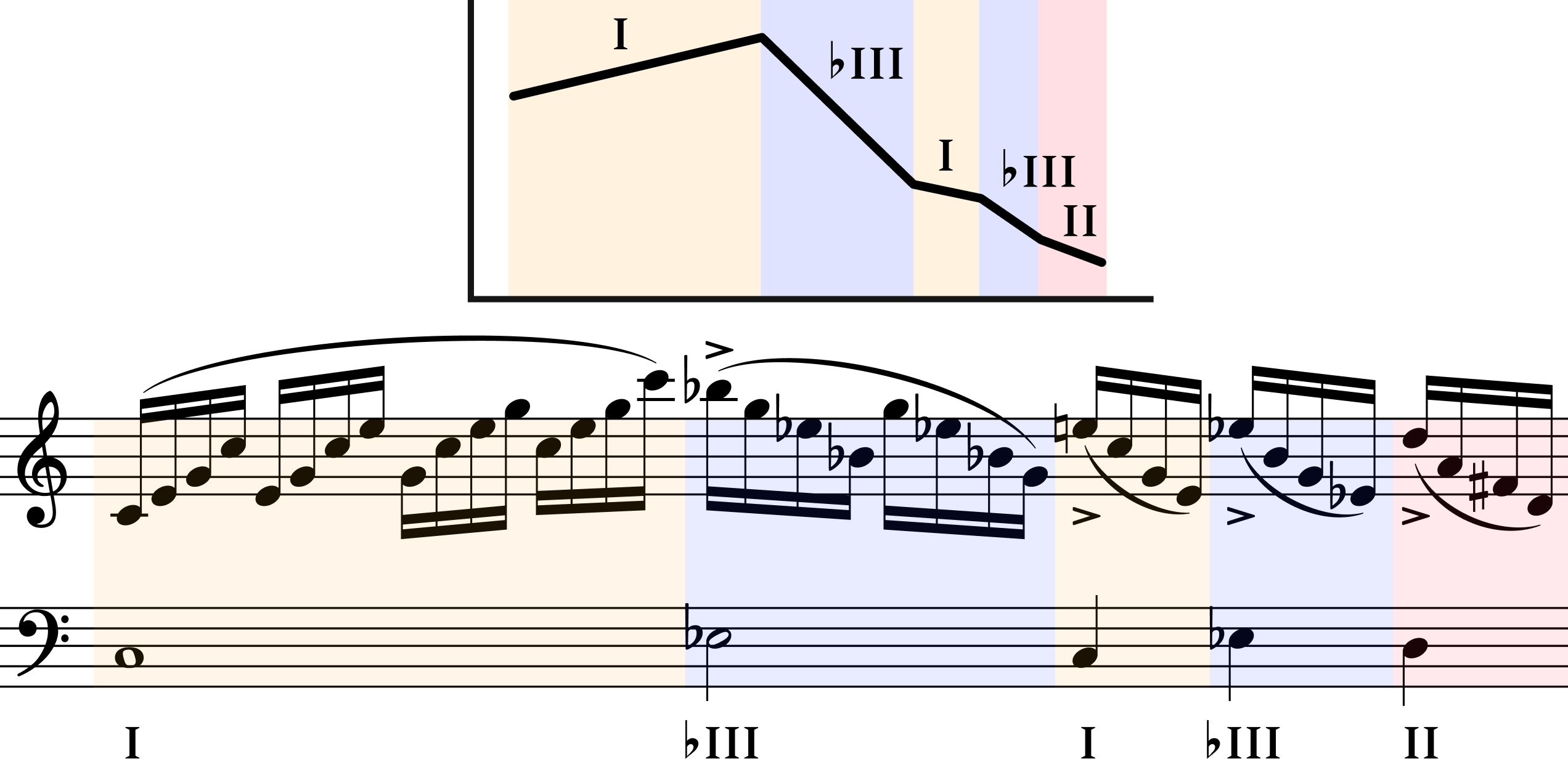}
 \caption{Musical notation for a line chart in major (\texttt{track 03}).}
 \label{fig:demo}
\end{figure}

At this juncture, we invite you to sit back, relax, and enjoy our classical data melodification soundtrack, available at \osflinkrepo. We guide your attention through our design space by changing its parameters in each of the tracks. In the first two tracks, we apply different musical colours to a melodified bar chart (categorical data) by changing the key chord progression of its ending cadence. Specifically, we play a perfect cadence (major to major key) to convey positive feelings in \texttt{track 01} and a deceptive cadence (major to minor key) to convey negative feelings in \texttt{track 02}. Next, we melodify a line chart (continuous quantitative data) and once again vary its musical colours through the tonality of chord progressions. We play the line chart using major chords in \texttt{track 03} (see \cref{fig:demo} for notation), the minor versions of these chords in \texttt{track 04}, and add a hint of chromaticism in \texttt{track 05}. In \texttt{track 06}, we switch the idiom to that of a pie chart (categorical data as proportions) and play it in major for a positive palette. We switch for a last time to the scatterplot idiom (quantitative data left unconnected) and adjust its parameters of data density, variance, and colour. \new{\texttt{Track 07} plays a scatterplot using notes from a major scale, using reverb from pedalling to convey low data density and a wide range of notes to convey high data variance. \texttt{Track 08} similarly plays a scatterplot in major, but uses no pedalling to convey high data density and a narrow range of notes to convey low variance. \texttt{Track 09} mixes parameters from the last two tracks with a twist, playing a scatterplot with high data density and high variance using notes from a chromatic scale.} We end with a bonus track of early experiments with the live music coding tool Sonic Pi\footnote{https://sonic-pi.net/}.
\\
\\
\texttt{
    \textbf{TRACKLIST:}
    \begin{itemize}
        \item \textbf{01 bar-positive} -- Bar chart ending on a perfect cadence (major-major key chord progression)
        \item \textbf{02 bar-negative} -- Bar chart ending on a deceptive cadence (major-minor key chord progression)
        \item \textbf{03 line-positive} -- Line chart in major
        \item \textbf{04 line-negative} -- Line chart in minor 
        \item \textbf{05 line-grey} -- Line chart in chromatic
        \item \textbf{06 pie-positive} -- Pie chart in major
        \item \textbf{07 scatter-loden-hivar} -- Scatterplot in major with high reverb and high range
        \item \textbf{08 scatter-hiden-lovar} -- Scatterplot in major with low reverb and low range
        \item \textbf{09 scatter-grey} -- Scatterplot in chromatic with low reverb and high range 
        \item \textbf{Bonus} -- Experimental sine waves in Sonic Pi
    \end{itemize} 
}

We further illustrate our melodification process by comparing the quantitative data melodified in \texttt{track 03} with its notation (see \cref{fig:notation}). We assign arbitrary values to the slope of the trend lines and reflected the slope in C-major chord progressions represented by roman numerals. \new{For example, a positive slope of one plays an ascending arpeggio in the key of the first note (I) of a C-major scale (still C-major), while a negative slope of two corresponds to a descending arpeggio in the key of the second note (II) of C-major (now D-major).} We colour-code the figure to visually show how changes in trend align with changes in chord progression. In audio, you can hear the start of these changes when the first note of the arpeggio is played with an accent to emphasize this transition.

\section{Coda}

What began as a fun thought experiment for aesthetic data sonification through musical rhetoric, with the requisite nod to IEEE VIS 2025 in Vienna as a center of classical music, turned for us into a deeper reflection on the mechanization of music and data, and how the mathematics of (classical) music may be modeled as emotion and mapped to data, following a design space heavily inspired by Blair et al.'s work~\cite{blair2024quantifying} on emotional response to design elements and data characteristics. 

We note that our proposed data melodification design space does not support precise and accurate value look-up. This issue of accuracy and precision is similarly a challenge for sonification generally~\cite{wang2022seeing}, and is not an issue we try to solve here. 

\textbf{Live-coding vs. tactile recording of data melodies---}We first explored the design space through live music coding via Sonic Pi, which allowed us to maintain an awareness for mathematical structure. Yet Sonic Pi introduced some limitations in terms of freeform exploration of encoding possibilities as we needed to stay within its programming framework. These limitations led us to return to the (literal) drawing board with our design space. By writing out the music notation on a whiteboard rather than in code, we were free to humanistically ``hear and draw data'' and ideate with less constraints. To test this conceptual design space, we played snippets of each visualization idiom on a physical piano, which afforded us spontaneous improvisation and revisions of our design space based on a musician's tacit knowledge of musical rhetoric, such as incorporating expressive techniques that cannot be easily replicated digitally, e.g., rubato from romantic music requires some experience to know when and how to fluctuate tempo to enhance emotional impact. This is, as one may expect, more difficult to encode in a program, although we are keen to continue exploring a programmatic realization of this design space in Sonic Pi or similar tools. 

\textbf{Flexibility of design space---}Through producing our tracklist, we were able to experiment with the flexibility of our design space. We found that musical genre can be easily applied to the different visualization idioms and immutable data characteristics, though with varying success when it comes to creating semantic associations (detailed in later paragraphs). Some musical parameters lend themselves to specific idioms, such as pitch/trend with arpeggios/line chart, or rhythm/density and range/variance with staccato notes/scatterplot. Other combinations are more questionable and require some creativity when composing with the design space. For instance, it is not sensible to use reverb for data density in a standard bar chart, \new{yet this becomes possible in a histogram that conveys information about the frequency and density of data points.}

\textbf{Balancing mathematics and aesthetics---}Balancing mathematical structures with audible aesthetics is challenging. We encountered this when applying our melodic design space to a scatterplot idiom, specifically when it came to mapping data values to the 12-note system of notes. While full use of the system affords more notes to use within a given value range and thus provided a more direct data-to-melody representation, the auditory experience echoes that of modern jazz (our apologies to modern jazz aficionados out there)---somewhat random, chaotic, best enjoyed on quiet repeat with a negroni. (Possibly) more pleasing instead is the use of only notes in a scale rather than all possible tones, but this choice requires expanding pitch range to cover all data point-to-note mappings, which can be deceptive in terms of the actual data variance conveyed by this design choice. 

A more successful attempt at merging mathematics and aesthetics leverages musical genre to (counter)balance the influence trends can have on a reader's emotion even without design intervention~\cite{blair2024quantifying}. We explored this aspect of our design space by playing line charts of quantitative data through major, minor, and chromatic sequences, but we also see this applying to bar charts of categorical data that have derived trends heard through ascending and descending chords. While we only demoed our design space on one instrument, we believe the design space can be easily expanded using multiple voices and instruments. This expansion adds to the rhetorical layer provided by musical genre, e.g., horns for a march or electric guitar for rock, that broadens the aesthetic space.

\textbf{Crafting unintentional semantics---}As we experimented with melodifying bar charts, we found that we needed to adjust the key of individual chords to avoid unintentional injection of rhetorical meaning. For instance, \texttt{track 02} features a bar chart played in C-major containing chords that, when played in succession, create major--minor key chord progressions producing negative feelings of uncertainty and suspense. This was a ``gotcha'' moment that we had not considered until we sat down and began improvising with this design space. \new{The fundamental characteristics of notes such as pitch and note density also have influence over emotional valence and activation, even more distinctly than keys and melodies~\cite{gabrielsson2016expression}. Since immutable data characteristics have been found to play a role in shaping emotions, e.g., positive trend inducing positive affect~\cite{blair2024quantifying}, we composed mappings in our design space to be similarly affective, e.g., trend mapped to pitch.} Future work could explore how best to combine musical devices to create semantic associations. While we informally tested the design space with a select few trained in classical music (who thus enjoyed the tracklist), we are curious to test melodification with those without this training.

This work bridged our personal hobbies (and for some of us, previous lives as musicians) into visualization design to re-imagine sonification practices for data. Inspired by the rhetorical devices of classical music, we compose a \textit{data melodification space} that we speculate induces less dread and/or tinnitus than sonification's mechanical tones. Through the process of ``hearing and sketching'' data melodies, we posit that true data melodification is a human experience because it requires tacit knowledge that is not easily translatable through programmable code. Sonification is more than the mapping of marks and channels to musical parameters; the rhetorical layer of data requires a balance of mathematical and emotional melodies best mediated by human intervention. \textit{\textbf{Fin.} }

\section*{Supplemental Materials}
\label{sec:supplemental_materials}

All supplemental materials are available on OSF at \osflinkrepo, released under a CC BY 4.0 license. The repository includes (1) audio clips of the design space and (2) manuscript figures.

\section*{Figure and Audio Credits}
\label{sec:figure_credits}
\cref{fig:galaxies}, sonification of the Mice Galaxies, is available under public domain. Audio clips linked in this manuscript, unless otherwise stated, are sourced from Wikimedia Commons. For \cref{fig:teaser}, \cref{fig:notation,fig:designspace,fig:demo}, and audio clips listed in the data melodification soundtrack, we as authors state that these are and remain under our own personal copyright, with permission to be used here. We also make them available under the \href{https://creativecommons.org/licenses/by/4.0/}{Creative Commons At\-tri\-bu\-tion 4.0 International (\ccLogo\,\ccAttribution\ \mbox{CC BY 4.0})} license and share them at \osflinkrepo.

\acknowledgments{
The authors thank producer Do(re)mi for providing jam sessions at the piano and discussions on music theory.}

\bibliographystyle{abbrv-doi}

\bibliography{template}
\end{document}